\documentclass[showpacs,amsmath,amssymb,twocolumn,prl,superscriptaddress]{revtex4-2}
\usepackage{amssymb}
\usepackage[dvips]{graphicx}
\usepackage{enumerate}
\usepackage{epsfig}
\usepackage{subfigure}
\usepackage{xcolor}
\usepackage[T1]{fontenc}
\usepackage{fullpage}
\usepackage{amsthm,amsfonts,amssymb,amscd,mathrsfs,xspace,framed}
\usepackage{amsmath}
\usepackage{color}
\usepackage{setspace}
\usepackage{url}
\usepackage{wrapfig}
\usepackage{enumitem}
\begin{document}

\title{\textit{In situ} control of integrated Kerr nonlinearity}

\author{Chaohan Cui}
\affiliation{James C. Wyant College of Optical Sciences, The University of Arizona, Tucson, Arizona 85721, USA}
\author{Liang Zhang}
\affiliation{James C. Wyant College of Optical Sciences, The University of Arizona, Tucson, Arizona 85721, USA}
\author{Linran Fan}
\email{lfan@optics.arizona.edu}
\affiliation{James C. Wyant College of Optical Sciences, The University of Arizona, Tucson, Arizona 85721, USA}

\maketitle

\textbf{
Kerr nonlinearity in nanophotonic cavities provides a versatile platform to explore fundamental physical sciences and develop novel photonic technologies~\cite{kippenberg2018dissipative, gaeta2019photonic, kues2019quantum}. This is driven by the precise dispersion control and significant field enhancement with nanoscale structures. Beyond dispersion and pump engineering, the direct control of Kerr nonlinearity can release the ultimate performance and functionality of photonic systems.
Here, we report the \textit{in situ} control of integrated Kerr nonlinearity through its interplay with the cascaded Pockels nonlinear process~\cite{schiek1993nonlinear,biaggio2001degenerate,kolleck2004cascaded}. Kerr nonlinearity is tuned over 10~dB dynamic range without modifying photonic structures. Fano resonance is observed with nonlinear spectrum, in contrast to standard linear transmission. This confirms the quantum interference between competing optical nonlinear pathways. Besides nonlinearity enhancement, we also demonstrate the novel capability to suppress the material intrinsic nonlinearity. Finally, we use the tunable nonlinearity to control the spectral brightness and coincidence-to-accidental ratio of single-photon generation. This work paves the way towards the complete nanoscale control of optical nonlinear processes at quantum level.
}


The dynamic control of optical properties plays an indispensable role in a great number of fields. The linear optical properties such as refractive index and absorption can be modified in nanophotonic circuits using a wide variety of phenomena including thermal~\cite{harris2014efficient}, electrical~\cite{wang2018nanophotonic,wang2018integrated}, mechanical~\cite{dong2010low}, and free-carrier effects~\cite{reed2010silicon}. It allows the modulation of optical fields with broad applications ranging from laser systems to optical communications. Recently, such linear control has also been used to demonstrate novel functions including the reconfigurable generation of topological photonic states~\cite{zhao2019non,zhang2020tunable}, the photonic acceleration of machine learning algorithms~\cite{shen2017deep,feldmann2019all}, and the realization of universal photonic quantum gates~\cite{carolan2015universal,wang2020integrated}.

\begin{figure}[htbp]
\centering
\includegraphics[width=\linewidth]{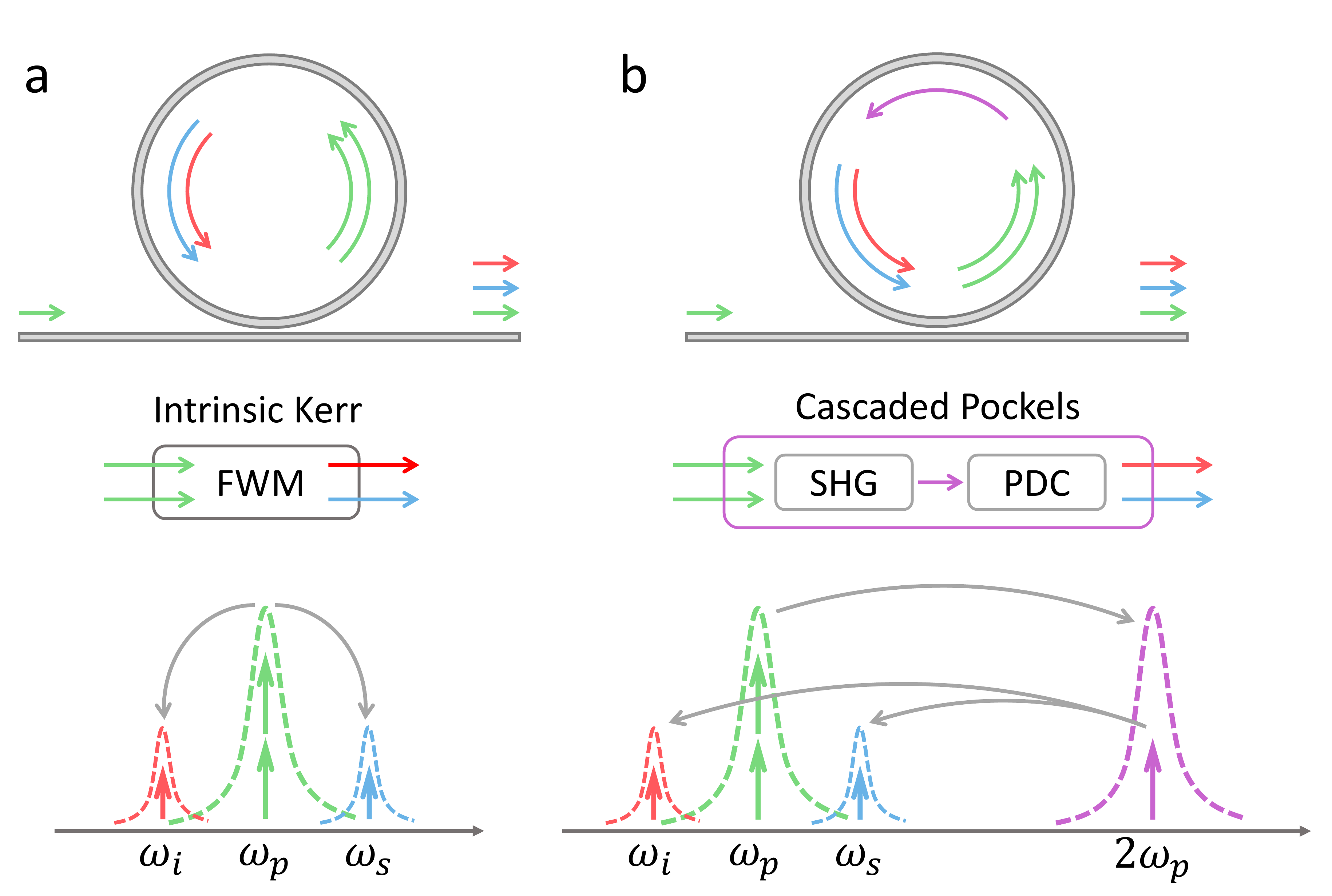}
\caption{\textbf{Effective Kerr nonlinearity with cascaded Pockels process.} \textbf{a}, Intrinsic Kerr process. Two pump photons are annihilated to generation one signal and one idler photon \textbf{b}, Cascaded Pockels process. Two pump photons are annihilated to generation one second-harmonic photon, which drives the parametric down-conversion to generate one signal and one idler photon.}
\label{fig:Fig1} 
\end{figure}

\begin{figure*}[htbp]
\centering
\includegraphics[width=\textwidth]{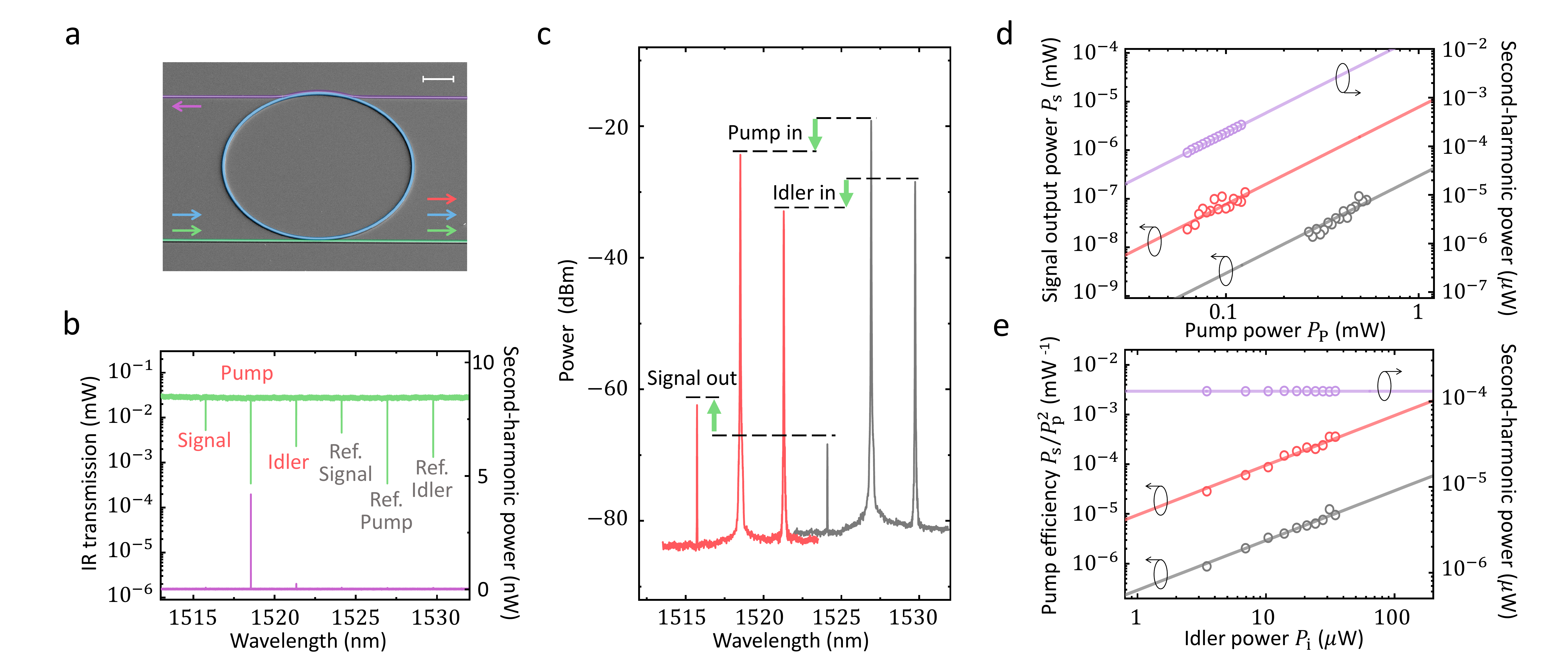}
\caption{
\textbf{Enhancement of Kerr nonlinearity.} \textbf{a}, Scanning electron microscope image of the fabricated aluminum nitride device with the ring cavity (blue), the bus waveguide for second-harmonic field (purple), and the bus waveguide for pump, idler, and signal fields (green). Scale bar 20~$\mu$m. \textbf{b}, Pump transmission (green) and second-harmonic (purple) spectrum. Two sets of modes for FWM, with (red) and without (gray) Pockels phase-matching, are labeled. \textbf{c}, The output optical spectrum of the stimulated FWM. \textbf{d}, The output signal (red and gray) and second-harmonic power (purple) with different pump power. \textbf{e}, The pump efficiency (red and gray) and second-harmonic power (purple) with different idler input power. Results in \textbf{c}, \textbf{d}, and \textbf{e} using the reference mode set and the mode set with Pockels phase-matching are presented in red and gray respectively. Experimental data (circles) are fitted with the theoretical model (solid lines).
}
\label{fig:Fig2} 
\end{figure*}

Similarly, the control of optical nonlinear properties will lead to significant scientific and technological advances. The nonlinearity enhancement into single-photon strong-coupling regime is essential to build deterministic quantum gates~\cite{milburn1989quantum, brod2016passive,heuck2020controlled}. The nonlinearity suppression is critical to fundamentally improve the sensitivity of optical sensors~\cite{iwatsuki1986kerr,liang2017resonant} and the capacity of optical communications~\cite{ellis2009approaching,le2017nonlinear}. The phase inversion of the nonlinear interaction can enable the generation of novel soliton states~\cite{xue2015mode, helgason2021dissipative}. Currently, optical nonlinear processes can only be controlled indirectly using intra-cavity photon number and dispersion. Therefore, the achievable nonlinearity
and functionality are limited to the material’s
intrinsic property.
In this Letter, we use a novel method to control Kerr nonlinearity beyond the material limit. By designing
its interference with cascaded Pockels processes~\cite{schiek1993nonlinear,biaggio2001degenerate,kolleck2004cascaded}, the
complete amplitude and phase control of Kerr nonlinearity can be realized. Besides the enhancement of Kerr nonlinearity~\cite{bosshard1999cascaded}, we also observe novel effects including the nonlinearity suppression and Fano resonances in the nonlinear regime. The effective Kerr nonlinearity is highly tunable over 10~dB dynamic range with fixed photonic structures. We further validate the control of Kerr nonlinearity in both classical and quantum regime through nonlinear frequency conversion and single-photon generation.


In the degenerate configuration of Kerr nonlinear processes, two pump photons are annihilated to generate one signal and one idler photon, or vice versa (Fig.~\ref{fig:Fig1}a). The nonlinearity is fixed by the material property and photonic structure. To modify the Kerr nonlinearity, we design a cascaded Pockels nonlinear process shown in Fig.~\ref{fig:Fig1}b. In this process, two pump photons first combine to generate one second-harmonic photon. Then the second-harmonic photon drives the parametric down-conversion process in the same photonic cavity to generate one signal and one idler photon. The effective interaction Hamiltonian for the cascaded Pockels nonlinear process is
\begin{equation}
    H_\mathrm{I} = -\frac{1}{2}\hbar (g_{c}\hat{b}_p^{\dagger2}\hat{b}_s\hat{b}_i+g_{c}^*\hat{b}_p^2\hat{b}_s^{\dagger}\hat{b}_i^{\dagger}),
    \label{H_22}
\end{equation}
with $\hat{b}_p$, $\hat{b}_s$, and $\hat{b}_i$ the annihilation operators of the pump, signal, and idler modes respectively (Supplementary Information section I). The effective cascaded Pockels nonlinear strength $g_c$ at single-photon level is defined as
\begin{equation}
    g_c=\frac{-i|g_2|^2}{i\Delta+\gamma/2},
\end{equation}
with $g_2$ the Pockels nonlinear strength, $\Delta$ the frequency detune between the pump and second-harmonic modes, and $\gamma$ the decay rate of the second-harmonic mode. This interaction Hamiltonian has the same form with intrinsic Kerr nonlinear processes. With highly efficient Pockels nonlinear processes, the effective Kerr nonlinearity can be significantly enhanced.

\begin{figure*}[htbp]
\centering
\includegraphics[width=\textwidth]{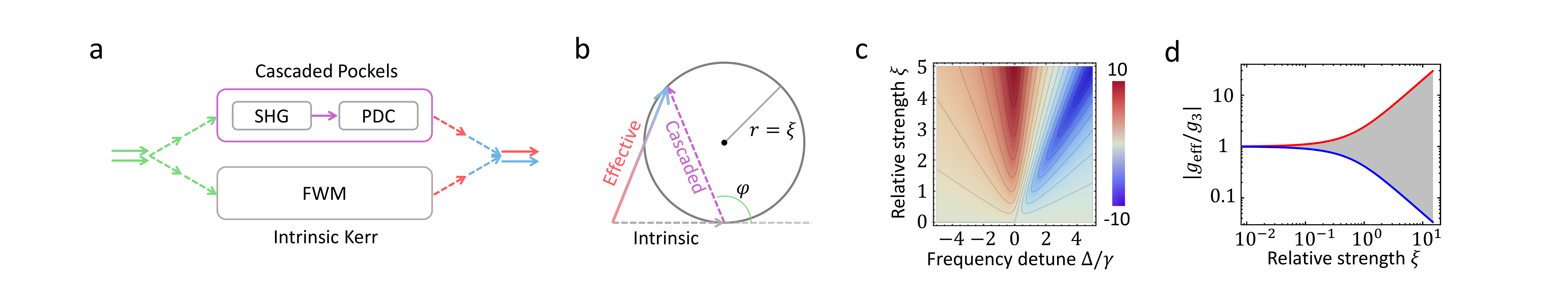}
\caption{\textbf{Quantum interference between nonlinear processes.} \textbf{a}, Schematic for the quantum interference between intrinsic Kerr and cascaded Pockels process. \textbf{b}, Vector diagram showing the coherent interaction between the intrinsic Kerr and cascaded Pockels process. Phase $\varphi$ can be controlled by the frequency detune $\Delta$. Intrinsic Kerr nonlinearity is normalized to 1, and the tuning range of the effective Kerr nonlinearity is bounded to the circle with radius $\xi$. \textbf{c}, Calculated effective Kerr nonlinearity $|g_{\rm eff}/g_3|$ in logarithm scale. \textbf{d}, Calculated maximum and minimum effective Kerr nonlinearity. The shaded area indicates the accessible range of the effective Kerr nonlinearity.}
\label{fig:Fig3} 
\end{figure*}

To implement the cascaded Pockels nonlinear process, we use a nanophotonic ring cavity made of aluminum nitride (Fig.~\ref{fig:Fig2}a). The phase-matching condition for the Pockels nonlinear process is satisfied using a high-order transverse-magnetic (TM) second-harmonic mode and a fundamental TM pump mode (Supplementary Information section II). Strong second-harmonic generation is observed with pump wavelength near 1518~nm (Fig.~\ref{fig:Fig2}b). The second-harmonic generation efficiency is measured as $\eta\approx1800\%/$W, leading to the estimated Pockels nonlinear strength $g_2\approx 2\pi\times81$~kHz (Supplementary Information section II). With the small decay rate of the second-harmonic mode ($\gamma\approx 2\pi \times 4.5$~GHz), we can expect the enhanced Kerr nonlinearity around $|g_{\rm c}| \approx 2\pi \times 3$~Hz at zero frequency detune, which is six-fold higher than the intrinsic value ($g_3\approx 2\pi \times 0.5$~Hz).

The enhancement of the Kerr nonlinearity is first verified with stimulated four-wave mixing (FWM) for frequency conversion in the classical domain. With strong pump ($P_p$), we measure the conversion from the idler input ($P_i$) to the signal output ($P_s$). We use two sets of modes from the same nanophotonic cavity to measure the intrinsic and enhanced Kerr single-photon nonlinearity respectively. The pump mode of the reference set is not phase-matched to the Pockels nonlinear process. Therefore, only the intrinsic Kerr nonlinearity contributes to the nonlinear frequency conversion. The pump mode of the other set is phase-matched to the Pockels nonlinear process. As a result, the enhanced Kerr nonlinearity dominates the nonlinear frequency conversion. The two sets of modes share similar quality factors and coupling conditions. Consequently, the influence of different cavity-enhancement factors and extraction efficiencies can be eliminated (Supplementary Information section II). The generation of the signal field is observed from the output optical spectrum with both mode sets (Fig.~\ref{fig:Fig2}c). Comparing the two mode sets, the signal output of the phase-matched set is significantly higher than the reference set even though smaller pump and idler inputs are used. To calibrate the enhancement factor, we fix the input idler power ($P_i$) and vary the input pump power ($P_p$). To achieve the same signal output ($P_s$), the cascaded Pockels nonlinear process uses 7.5~dB less pump power than the intrinsic Kerr nonlinearity (Fig.~\ref{fig:Fig2}d). This agrees well with the difference between the enhanced and intrinsic Kerr single-photon nonlinearity ($|g_c/g_3|\approx 6$).The critical role of the Pockels nonlinear process is further confirmed by the increased second-harmonic generation with respect to the pump power (Fig.~\ref{fig:Fig2}d). The same conclusion is obtained from the measurement with different input idler power ($P_i$). With the same pump power, the cascaded Pockels process can generate approximately 15~dB stronger signal output. The second-harmonic generation remains constant during the process, which confirms that the second-harmonic field is generated from the pump.

\begin{figure*}[hbtp]
\centering
\includegraphics[width=\textwidth]{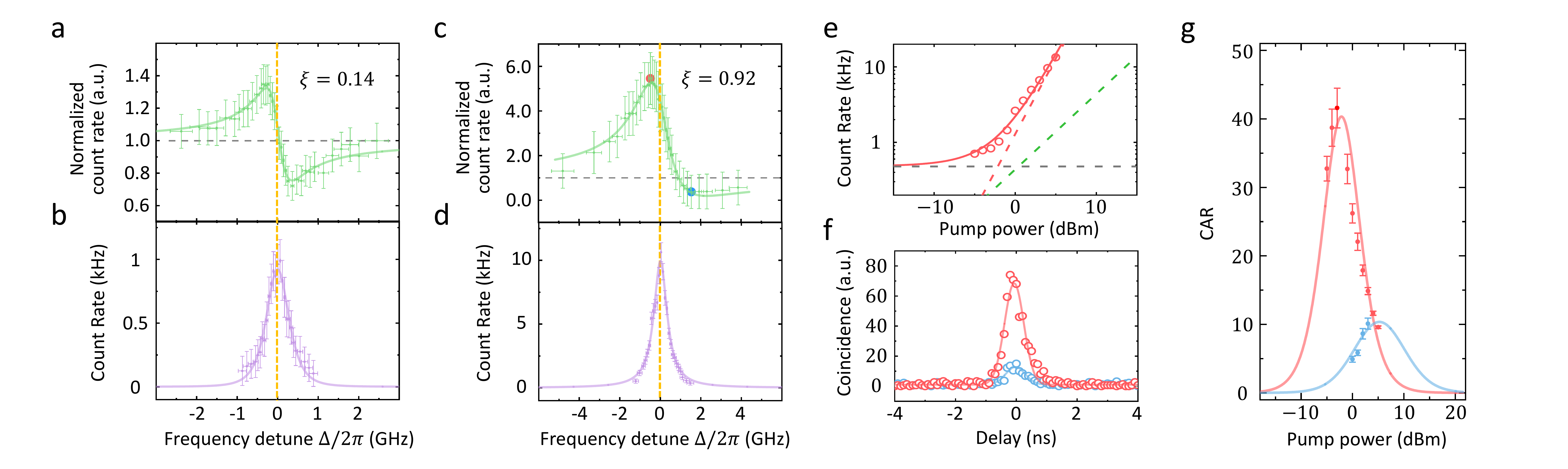}
\caption{\textbf{Continuous tuning of Kerr single-photon nonlinearity in quantum regime.} \textbf{a}, \textbf{c}, Single-photon generation rate from spontaneous FWM using effective Kerr nonlinearity with $\xi=0.14$ and $\xi=0.92$ respectively. The data is normalized to the count rate with the intrinsic Kerr nonlinearity (gray dashed line). \textbf{b}, \textbf{d}, Single-photon generation rate from spontaneous parametric down-conversion with $\xi=0.14$ and $\xi=0.92$ respectively. \textbf{e}, Single-photon generation rate (red circle) with different pump power. Calculated background dark count (black dashed), pump-induced noise (green dashed), signal photon rate (red dashed), and total photon rate (red solid) are plotted. All photon counting data are measured from signal resonances. \textbf{f}, Signal-idler coincidence count with frequency detune $\Delta=-2\pi \times 0.7$~GHz (red point) and $\Delta=2\pi \times 1.5$~GHz (blue point). Pump power is 1 mW. The accidental coincidence outside the coherent window is normalized to 1. \textbf{g}, CAR under different pump power with frequency detune $\Delta=-2\pi \times 0.7$~GHz (red) and $\Delta=2\pi \times 1.5$~GHz (blue). Circle, measured CAR from signal-idler coincidence. Line, calculated CAR from calibrated signal and noise photon rates.
}
\label{fig:Fig4} 
\end{figure*}

The coherent interaction between multiple nonlinear processes can be used to tune the overall nonlinear response. This enables the complete control over the amplitude and phase of the Kerr single-photon nonlinearity. 
Taking both the intrinsic Kerr and cascaded Pockels nonlinear processes into account (Fig.~\ref{fig:Fig3}a), the effective Kerr nonlinearity can be written as 
\begin{align}
    g_{\rm eff} = g_3+g_c = g_3-\frac{ig_2^2}{i\Delta+\gamma/2}
\end{align}
The relative phase $\varphi$ between the two nonlinear processes can be controlled by the frequency detune~$\Delta$ (Fig.~\ref{fig:Fig3}b). Both constructive and destructive interference can be realized, leading to the enhancement and suppression of the Kerr nonlinearity respectively (Fig.~\ref{fig:Fig3}c). Here, we define the relative strength between the two nonlinear processes as
\begin{align}
\xi=|g_2|^2/(g_3\gamma).
\end{align}
The maximum and minimum effective single-photon nonlinearity $|g_{\rm max/min}|=g_3\left(\frac{\sqrt{1+\xi^2}\pm\xi}{\sqrt{1+\xi^2}\mp\xi}\right)^{1/2}$ can be obtained at frequency detunes $\Delta=\frac{\gamma }{2}(\xi\mp\sqrt{1+\xi^2})$. Therefore, larger tuning dynamic range can be achieved with more efficient Pockels nonlinear processes (Fig.~\ref{fig:Fig3}d).

To verify the quantum interference between the cascaded Pockels and the intrinsic Kerr nonlinear processes, we perform single-photon generation with spontaneous FWM. The frequency detune $\Delta$ is precisely controlled through the device temperature (Supplementary Information section II). The pump light is tuned in resonance with the pump mode to ensure that the intra-cavity pump photon number remains constant. The single-photon generation rate of the signal channel (proportional to $|g_{\rm eff}|^2$) is recorded to infer the effective Kerr single-photon nonlinearity. We use two different Pockels nonlinear strengths corresponding to $\xi=0.14$ and $\xi=0.92$ respectively by selecting two different phase-matching modes. The single-photon count rate shows Fano lineshape in both cases (Fig.~\ref{fig:Fig4}a and c), indicating the coherent interaction between two competing optical nonlinear processes. Larger tuning dynamic range of the Kerr nonlinearity is achieved with the stronger Pockels nonlinear strength ($\pm0.6$~dB with $\xi=0.14$ vs. $\pm3.6$~dB with $\xi=0.92$), which matches our theoretical model (Fig.~\ref{fig:Fig3}c). As reference, we also measure the single-photon generation rate with parametric down-conversion by pumping a visible laser into the cavity under different frequency detune (Fig.~\ref{fig:Fig4}b and d), which shows Lorentzian lineshape. This confirms the quantum interference between the broad-band intrinsic Kerr process and the narrow-band cascaded Pockels nonlinear process. It is noteworthy that Fano resonances in our experiment can only be observed with nonlinear spectrum, as the interference happens between highly nonlinear processes. This is in contrast to Fano resonances introduced by linear coupling between resonators, where linear transmission is sufficient to observe the asymmetric lineshape~\cite{limonov2017fano}.

Finally, we demonstrate the control of the coincidence-to-accidental ratio (CAR) of the single-photon generation by changing the Kerr single-photon nonlinearity. CAR is a critical parameter to characterize the quality of single photon sources. In the low pump regime, background noise and leaking pump dominate the total photon count. Therefore, CAR can be improved by increasing the pump power. In the high pump regime, multi-photon generation becomes the major noise source. As a result, CAR drops with increased pump power. We first set the frequency detune $\Delta = 2\pi \times 1.5$~GHz with $\xi=0.92$ to achieve the maximum nonlinearity suppression (blue point in Fig.~\ref{fig:Fig4}c). We measure the coincidence count between signal and idler photons to extract the dependence of CAR on the pump power (Fig.~\ref{fig:Fig4}f and g). The maximum CAR of $9\pm1$ is achieved using 3~dBm pump power. Next, we set the frequency detune $\Delta = -2\pi \times 0.7$~GHz to realize the maximum nonlinearity enhancement (red points in Fig.~\ref{fig:Fig4}c). The maximum achievable CAR is $47\pm2$ with -2~dBm pump power, corresponding to more than five-fold improvement. We also calibrated photon rates of the background noise, pump-induced noise, and parametric photon pairs (Fig.~\ref{fig:Fig4}e) to directly calculate CAR, which agrees with the result obtained from the coincidence measurement (Fig.~\ref{fig:Fig4}f) (Supplementary Information III).

We have demonstrated the \textit{in situ} control of local Kerr nonlinearity by designing the quantum interference between intrinsic Kerr and cascaded Pockels processes. Nonlinear efficiency and novel functionalities beyond material limits are realized. Further improvement can be readily achieved using nanophotonic materials with larger Pockels nonlinearity~\cite{elshaari2020hybrid,lu2020toward,zhu2021integrated} and low loss design~\cite{desiatov2019ultra}. It can benefit a wide range of fields requiring strong Kerr nonlinearity, including all-photonic signal processing~\cite{foster2008silicon}, nonlinear frequency conversion~\cite{lu2019efficient}, and optical frequency comb~\cite{miller2014chip,jung2014electrical,pu2016efficient,stern2018battery,chang2020ultra}. More intriguingly, the demonstration of nonlinearity suppression and phase control can open unique prospects for the study of nonlinear physics and photonic technologies in both classical and quantum regime. 


\textbf{Methods}

\textit{Device fabrication.} Devices are fabricated from 1-$\mu$m AlN grown on sapphire substrates by MOCVD. FOx-16 resist is used for patterning photonic circuits with electron-beam lithography. After development with TMAH, plasma etching with ${\rm Cl}_2$/${\rm BCl}_3$/Ar is used to transfer the pattern to the AlN layer. Finally, ${\rm SiO}_2$ cladding is deposited by plasma-enhanced chemical vapor deposition (PECVD).

\textit{CAR measurement.} When calculating CAR, the coincidence count is averaged within the coincidence peak ($\pm$ 1 ns) at zero delay. The accidental count is estimated by averaging the coincidence count with large time delay outside the peak. 

\textbf{Acknowledgments}
CC, LZ, and LF acknowledge the support from U.S. Department of Energy, Office of Advanced Scientific Computing Research, (Field Work Proposal ERKJ355); Office of Naval Research (N00014-19-1-2190); National Science Foundation (ECCS-1842559). Device fabrication is performanced in the OSC cleanrooms at the University of Arizona, and the cleanroom of Arizona State University. Superconducting nanowire single-photon detectors are supported by NSF MRI INQUIRE.

\textbf{Data availability}
The data that support the findings of this study are available from the corresponding author upon reasonable request

\bibliography{Ref}

\end{document}